\documentclass[paper=a4, fontsize=11pt]{scrartcl}
\usepackage[T1]{fontenc}
\usepackage{fourier}

\usepackage[english]{babel}								
\usepackage[protrusion=true,expansion=true]{microtype}	
\usepackage{amsmath,amsfonts,amsthm}
\usepackage[pdftex]{graphicx}	
\usepackage{url}
\usepackage{siunitx}

\usepackage[pdfencoding=auto]{hyperref}
\usepackage{subcaption}
\usepackage[sorting=none,maxnames=6]{biblatex}
\usepackage{sectsty}
\allsectionsfont{ \normalfont} 

\usepackage{fancyhdr}
\numberwithin{equation}{section}		
\numberwithin{figure}{section}			
\numberwithin{table}{section}				

\newcommand{\horrule}[1]{\rule{\linewidth}{#1}} 	

\title{%
    \texorpdfstring{
		\usefont{OT1}{bch}{b}{n}
		\normalfont \normalsize \textsc{GSI Accelerator Physics Department} \\ [25pt]
		\horrule{0.5pt} \\[0.4cm]
		\huge Impact of Field Errors of Quadrupole Units on Beam Losses at SIS100 Injection \\
		\horrule{2pt} \\[0.5cm]
	}{%
	   Impact of Field Errors of Quadrupole Units on the Space Charge induced Beam Loss at SIS100 Injection%
	}%
}
\newcommand{\aujordhui}{22 September 2021}
\author{%
    \texorpdfstring{
		\normalfont 								\normalsize
        Adrian Oeftiger\footnote{a.oeftiger@gsi.de} \\[-3pt]
        \normalsize\textit{ GSI Helmholtzzentrum f\"ur Schwerionenforschung GmbH, Darmstadt, Germany} \\[-3pt]
        \normalsize
        \aujordhui \\[1em]
        \normalsize
        GSI Report: GSI-2021-00450
    }{%
        Adrian Oeftiger%
    }%
}
\date{\texorpdfstring{}{\aujordhui}}

\begin{document}
\maketitle

\begin{abstract}
    \emph{Abstract:} This report discusses heavy-ion beam losses due to the nominal transverse space charge and the field imperfections in SIS100 during accumulation based on recent cold measurements of the first 10 series quadrupole units. They are compared to the (stamped) First-of-Series quadrupole unit \#2.
\end{abstract}

\section{Overview}

The field quality of the steering and focusing magnet units in the SIS100 have a strong impact on the maximally achievable intensities.
As the SIS100 accumulates eight bunches over four injections from SIS18 during heavy-ion beam accumulation, the first injected bunches circulate at the rigidity of \SI{18}{\tesla\meter} during a storage time of \SI{1}{\second} before the ramp starts.
For the reference heavy-ion beam type Uranium-238 (charge state $28+$) this duration translates to 160'000 turns.
During this time the bunches can be subject to emittance growth and subsequent beam loss due to resonances driven by magnetic field imperfections. 
Due to an effective space charge tune spread of up to $\Delta Q^{SC}_y=-0.3$ the resonances cover larger areas in the tune diagram.

\section{Magnet Field Error Model}

For this study we focused on the SIS100 heavy-ion fast extraction optics working point, located in the tune quadrant $18.5<Q_{x,y}<19$.
To obtain tune figures in this quadrant, the $L\approx\SI{1.3}{\meter}$ long superferric quadrupole magnets need to be powered at an integral strength of around $K_1L=\SI{0.28}{\per\meter}$. 
The thus required focusing strength $K_1$ then determines the quadrupole gradient $\partial B_y/\partial x$ at a given beam rigidity $B\rho$ via
\begin{equation}
    K_1 = \frac{1}{B\rho}\,\frac{\partial B_y}{\partial x} \quad .
\end{equation}
At SIS100 injection with $B\rho=\SI{18}{\tesla\meter}$ the required quadrupole gradient hence amounts to
\begin{equation}
    \frac{\partial B_y}{\partial x} = \SI{18}{\tesla\meter}~ \frac{0.28}{\SI{1.3}{\meter^2}} = 3.9\, \frac{\si{\tesla}}{\si{\meter}} \quad .
\end{equation}
In the bench measurements, the most suitable setup reaching a comparable gradient is at a current of $I=\SI{900}{\ampere}$.
The quadrupole field error models in the following are based on the measured nonlinear multipole components at this current.
On the linear order, a normal gradient error of rms amplitude $b_2=\SI{5}{units}$ (\SI{1}{unit} conventionally refers to \SI{1d-4}{}) is assumed for all magnets \cite{keigraderror} while the skew component is neglected, $a_2=0$.

The field error model for the dipole magnet units is described in Ref.\ \cite[Section 2]{newslettersis100} and based on the bench measurements carried out by F.\ Kaether et al.

\subsection{First-of-Series Quadrupole Unit \#2}
The multipole measurements of the second First-of-Series (FoS) quadrupole unit (the yoke of which has been stamped and not milled after assembly) \cite{fos} are used to create a field error model for simulations henceforth referred to as ``FoS 2'' model.
The nonlinear part has been described in detail in Ref.\ \cite[Section 2]{newslettersis100}.
Beam simulations demonstrate that the multipole errors beyond dodecapolar order $b_6$ (with figures around or below \SI{0.5}{units}) have no effect on the losses with space charge during accumulation of heavy ions.

\begin{figure}[htbp]
    \begin{subfigure}[t]{0.49\linewidth}
        \includegraphics[width=\linewidth]{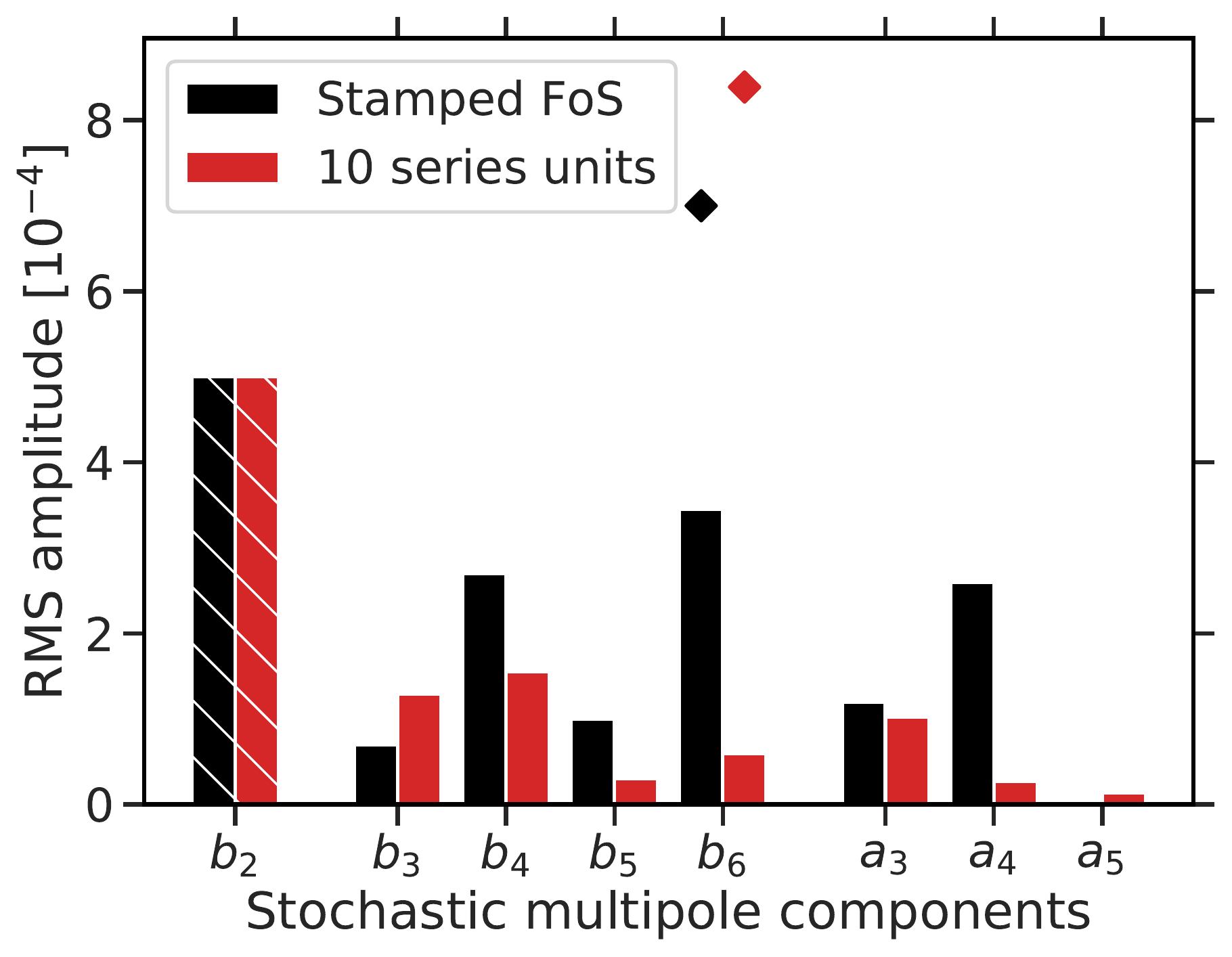}
        \caption{First-of-Series \#2 vs.\ ensemble (based on 10 series magnets), the systematic component in $b_6$ is indicated by diamond markers. The assumed $b_2$ component appears hatched.}
        \label{fig: err model ensemble vs fos}
    \end{subfigure}
    \hfill
    \begin{subfigure}[t]{0.49\linewidth}
        \includegraphics[width=\linewidth]{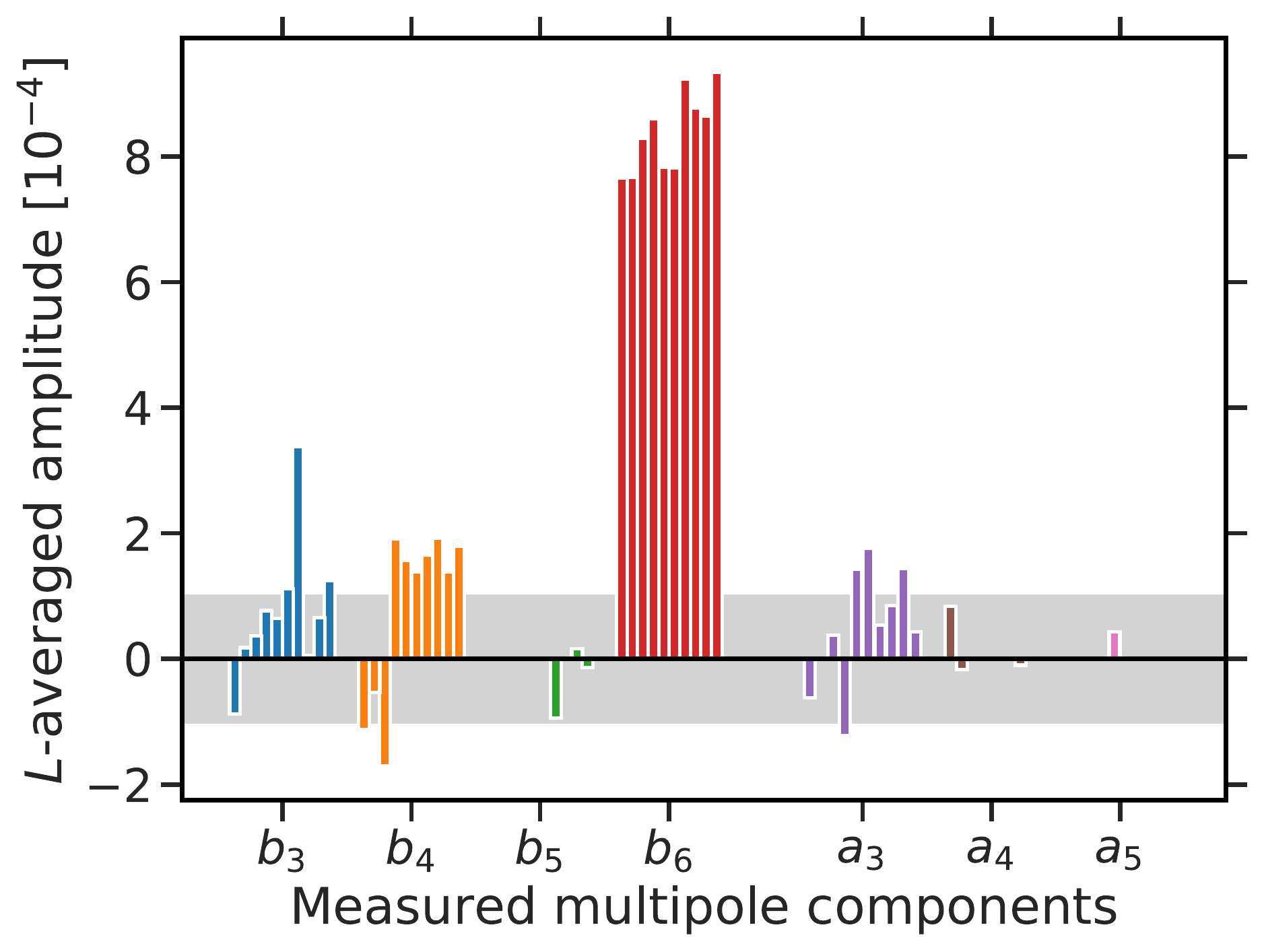}
        \caption{10 series magnets, one bar marks the amplitude averaged along the magnet length $L$ for the respective component. The grey area indicates an interval of \SI{1}{unit} radius.}
        \label{fig: err model 10 new}
    \end{subfigure}
    \caption{Rms amplitudes of multipole components for stochastic field error models in simulations ($b_2$ is assumed to be \SI{5d-4}{}, all others are based on measurements).}
\end{figure}

The black bars in Fig.\ \ref{fig: err model ensemble vs fos} indicate the most relevant multipole components in the FoS 2 model\footnote{Based on insights from magnet simulations, some figures from the FoS 2 measurements have been considered lower than the expected average. 
Thus, the following values have been slightly increased in the FoS 2 field error model to remain on the conservative side, viz.\ $b_3=0.7$ (not $0.6$), $b_5=1.0$ (not $0.9$), $a_3=1.2$ (not $0.5$) and $a_6=1.0$ (not $0.8$).}.
The model contains no systematic components besides $b_6$, which is assumed to amount to the measured \SI{6.9}{units}.
The corresponding stochastic rms amplitude is taken as half the measured value, $b_6=\SI{3.45}{units}$.

\subsection{Ensemble of Series Quadrupole Units}

Steady series production of the quadrupole units has commenced and cold bench measurements of the first 10 units have provided a set of field imperfections\footnote{The measurement protocols of the 10 cold quadrupole units bear the CIDs 02000090630, 02000090647, 02000090654, 02000090661, 02000090678, 02000091330, 02000091354, 02000091361, 02000091378, 02000091385 in the respectively tagged reports prepared by JINR Institute, Dubna.}. 
Figure \ref{fig: err model 10 new} displays the measured multipole components, the bars each stand for one quadrupole magnet (sorted by the respective report CID).
In comparison to the FoS 2 model from Fig.\ \ref{fig: err model ensemble vs fos}, the normal $b_6$ component has slightly grown to values fluctuating around a mean of \SI{8.4}{units} while the normal $b_4$ component has approximately halved and the skew $a_4$ component is strongly reduced for all 10 series quadrupole units.
The grey area marks an $\pm$\SI{1}{unit} interval to guide the eye.

The field error model for the beam dynamics simulations, which is based on these 10 series unit measurements, is henceforth referred to as the \emph{ensemble model}.
The stochastic multipole components determine the resonance driving terms, they are estimated from the spread of individual magnets around the systematic figure.
The stochastic components of the ensemble model are exhibited as red bars in Fig.\ \ref{fig: err model ensemble vs fos}.
The average figures of all magnets for each component give a hint on the corresponding systematic component.
Except for the aforementioned dodecapolar $\langle b_6 \rangle = \SI{8.4}{units}$, all average figures lie below one unit.
For a conservative estimate of the resonance driving terms in the simulations, the systematic components apart from $b_6$ are considered to be zero.
The root mean square $\sqrt{\sum_{i=1}^{10} [(b_n)_i - \langle b_n\rangle]^2}$ then yields the stochastic rms amplitude.

\subsection{Comparison FoS \#2 and Ensemble Models}

Figure \ref{fig: err model ensemble vs fos} depicts the stochastic rms amplitude figures of the two error models.
In comparison to the FoS 2 error model, the ensemble model features
\begin{itemize} \setlength{\itemsep}{0.2em}
    \item a doubled normal sextupolar component ($\approx$ factor 1.8),
    \item a halved normal octupolar component ($\approx$ factor 0.6),
    \item strongly reduced normal dodecapolar and skew octupolar components.
\end{itemize}

\clearpage
\section{Setup of Beam Dynamics Simulations}
The full beam dynamics simulations of a single bunch in SIS100 with nonlinear space charge, nonlinear thin-lens tracking and nonlinear RF bucket are carried out using the open source CERN code SixTrackLib \cite{sixtracklib, sixtrack, sixtracklibgithub}.
The full SIS100 lattice \cite{syslatticerepo} is simulated based on the heavy-ion fast extraction tune settings, where the working point is scanned in $0.01$ intervals across the tunes $18.55\leq Q_{x,y}<19$.
All beam and machine parameters are listed in Table \ref{tab: params}.

\begin{table}[htbp]
    \centering
    \caption{Parameters for Uranium-238 Beam Production in SIS100.}
    \renewcommand{\arraystretch}{1.3}
    \label{tab: params}
    \begin{tabular}{r|l}
        \textbf{Parameter} & \textbf{Value} \\ \hline
        Horizontal normalised rms emittance $\epsilon_x$ & \SI{5.9}{\milli\meter.mrad} \\
        Horizontal geometric KV emittance & \SI{35}{\milli\meter.mrad} \\
        Vertical normalised rms emittance $\epsilon_y$ & \SI{2.5}{\milli\meter.mrad} \\
        Vertical geometric KV emittance & \SI{15}{\milli\meter.mrad} \\
        Rms bunch length $\sigma_z$ & \SI{13.2}{\meter} \\
        Rms momentum deviation $\sigma_{\Delta p/p_0}$ & \SI{0.44d-3}{} \\
        Bunch intensity $N$ of U${}^{28+}_{238}$ & \SI{0.625d11}{} \\
        Max.\ space charge tune shift $\Delta Q^{SC}_y$ & $-0.30$ \\
        Chromatic tune spread $Q'_{x,y}\sigma_{\Delta p/p_0}$ & $0.02$ \\
        RF voltage (single-harmonic) $V_{RF}$ & \SI{58.2}{\kilo\volt} \\
        Harmonic $h$ & $10$ \\
        Kinetic energy & $E_{kin}=\SI{200}{\mega\electronvolt}/$u \\
        Relativistic $\beta$ factor & 0.568 \\
        Revolution frequency $f_{rev}$ & \SI{157}{\kilo\hertz} \\
    \end{tabular}
\end{table}

The simulations sample the bunch with 1000 macro-particles using an optics-matched 6D Gaussian phase space distribution.
This bunch is tracked through the synchrotron elements and space charge nodes, 501 of which are placed in intervals of slightly varying length along the ring.
Using more space charge nodes or more macro-particles does not significantly alter the simulation results.
Space charge is modelled as fixed (i.e.\ non-adaptive) frozen 3D Gaussian field maps which remain constant throughout the simulation. 
The transverse beam size used in the field maps at each space charge node follows the computed local $\beta$- and dispersion functions and is based on the initial transverse emittances.
The transverse nonlinear space charge force is modulated with the longitudinal Gaussian beam profile. 
Evaluating the initial tune footprint in the simulation gives a maximum space charge tune shift from the bare working point in the vertical plane of $\Delta Q^{SC}_y=-0.3$ for the particles in the centre of the bunch.

Without the multipole errors, the considered SIS100 lattice maintains an unbroken six-fold symmetry with dipole and quadrupole magnet units each sliced into 9 thin-lens slices. 
The $\beta$-beating effect of the longer radiation-hardened quadrupole units is assumed to be fully corrected.
To relate all non-systematic resonances to the used field error model, the SIS100 model here replaces the radiation-hardened quadrupole units by their shorter superconducting counterpart.
All dipole and quadrupole magnets around the ring are assigned multipole error components following a Gaussian distributed random sequence of stochastic rms amplitude (cut at 2 rms figures) as described in the previous section. 
All 9 thin slices of each magnet are assigned identical error figures.
The random number sequence is ensured to be identical across tunes and even separate simulation scenarios which are compared in this report.
The results based on the chosen random seed show a conservative resonance behaviour when comparing among 10 different random seeds.

Misalignment errors are not specifically taken into account in the simulations carried out for this report.
They (i.) result in feed-down for the multipole components (only minor impact for reasonable rms orbit deviation $\mathcal{O}(\SI{1}{\milli\meter})$) and (ii.) lead to diverging closed orbit distortion close to integer tunes (potentially dangerous impact for large rms orbit deviation, as beam loss can increase below the integer $Q_x=19$ due to orbit deviations which move the beam into the aperture).

\subsection{Interpretation of Results from Fixed Frozen vs.\ Self-consistent Particle-in-cell Space Charge Models} 

It is important to note that the assumption of a fixed frozen space charge model leads to a conservative estimate of the size of loss-free working point areas: working points without beam loss in the fixed frozen case normally remain loss-free in realistic simulations employing self-consistent 3D particle-in-cell (PIC) algorithms.
On the other hand, working points of low but finite beam loss can potentially remain entirely free of beam loss in the PIC case.
Finally, significant beam loss figures predicted by fixed frozen models tend to underestimate the more realistic figures from PIC, so that absolute finite beam loss figures in the following tune diagram plots are not necessarily meaningful (beyond qualitative indication of potentially dangerous resonance lines).
All in all, the assumption of fixed frozen field maps for space charge enables fast and conservative prediction of loss-free working point areas, the bounding contours of which might become larger (but usually not smaller) when predicted by computationally expensive high-resolution PIC simulations.

\clearpage
\section{Simulation Results}

\begin{figure}[htbp]
    \begin{subfigure}{0.49\linewidth}
        \includegraphics[width=\linewidth]{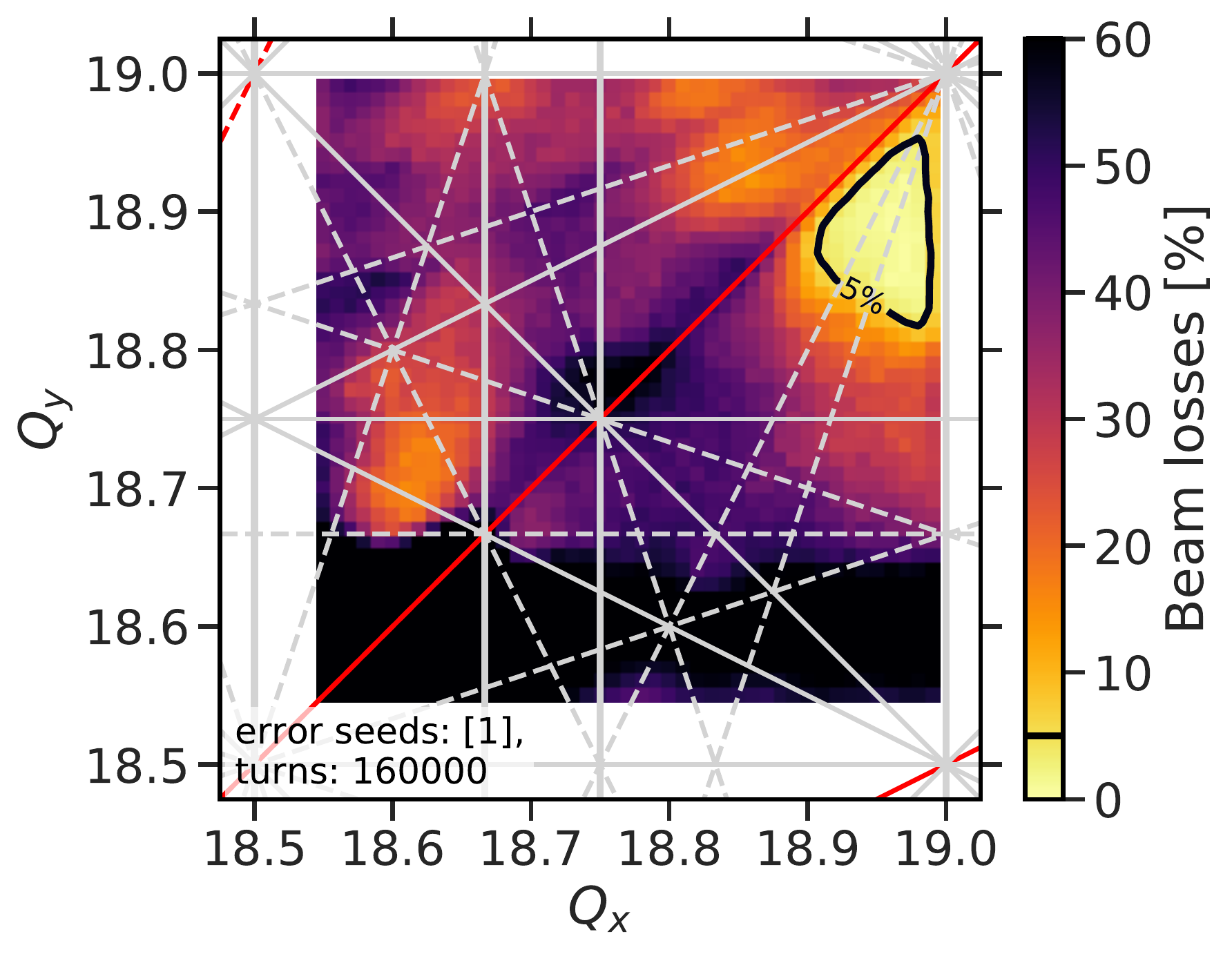}
        \caption{FoS 2 field error model.}
        \label{fig: loss fos}
    \end{subfigure}
    \hfill
    \begin{subfigure}{0.49\linewidth}
        \includegraphics[width=\linewidth]{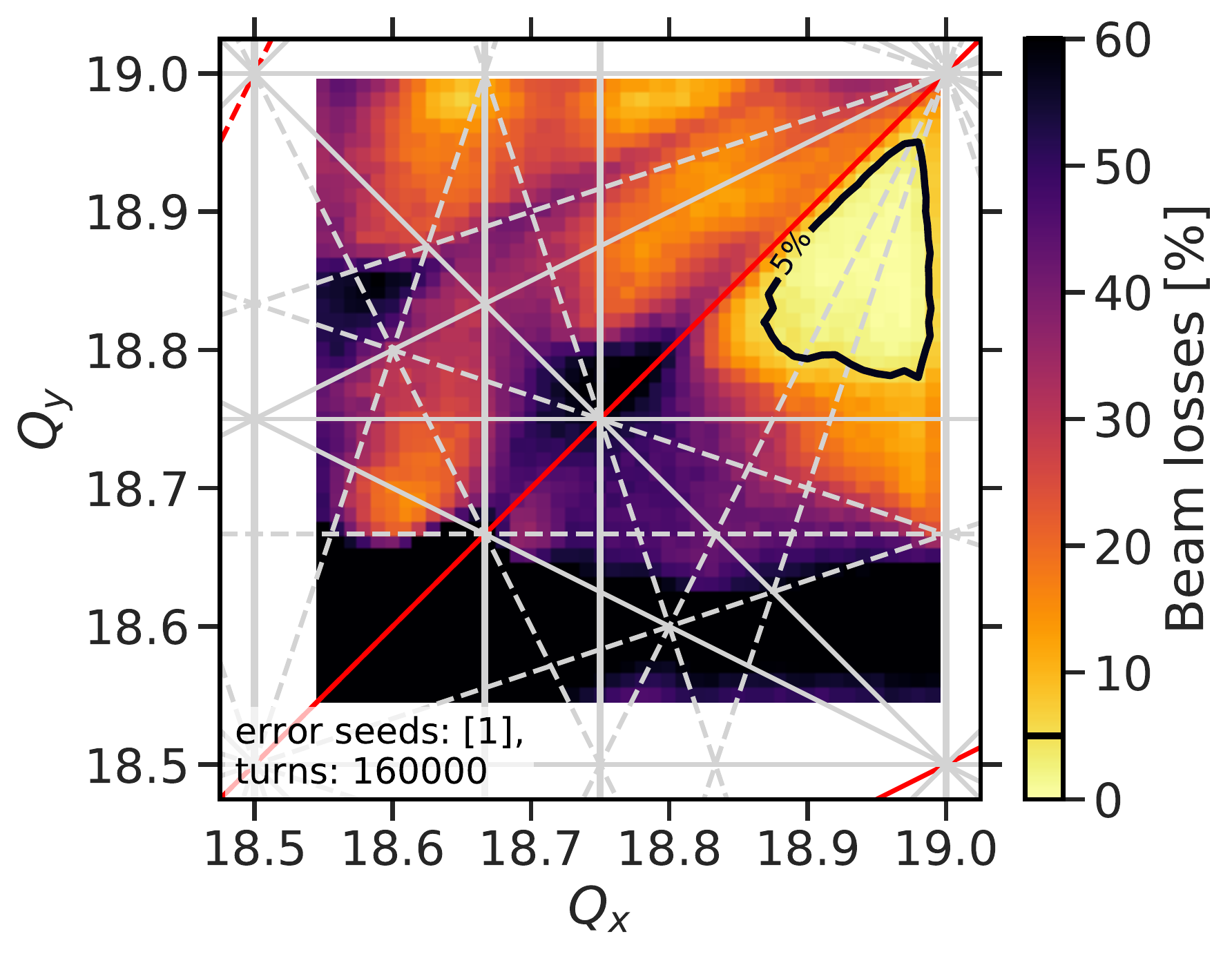}
        \caption{Ensemble field error model.}
        \label{fig: loss ensemble}
    \end{subfigure}
    \caption{Beam loss tune diagrams during full \SI{1}{\second} injection plateau for different quadrupole nonlinear field error models. Resonance lines up to octupolar order are displayed (solid: normal, dashed: skew, grey: non-systematic, red: systematic).}
\end{figure}

For the FoS 2 field error model, the beam loss figures in the considered part of the tune diagram are presented in Fig.\ \ref{fig: loss fos}.
A good working point area is located around $Q_x,Q_y=(18.95, 18.87)$ indicated by yellow colour and enclosed by the black contour, inside of which the predicted beam loss across the injection plateau remains below $5\%$.

The ensemble model based on the 10 series quadrupole units results in the beam losses presented in the tune diagram in Fig.\ \ref{fig: loss ensemble}.
The available low-loss tune area is significantly enlarged towards the bottom in comparison to the FoS 2 model scenario. 

\subsection{Comments on Results from FoS 2 Field Error Model}

The FoS 2 implied low-loss area is limited to the top by the coupling line\footnote{The Montague (coupling) resonance $2Q_y-2Q_x$ is the reason for the limit on top and it is intrinsically driven by the nonlinearities of the space charge fields:
for working points in the vicinity of the coupling line $Q_x=Q_y$, the $\approx 2.5\times$ larger horizontal emittance is exchanged with the vertical plane where the aperture is much more restrictive.
Therefore, the Montague resonance leads to strong beam loss around the coupling line, which significantly exceeds the beam loss contribution from skew quadrupole components $a_2$ within a reasonable range of tens of units ($a_2$ are not taken into account in the error models here though).}
$Q_x=Q_y$ and to the right by the integer resonance $Q_x=19$.
Both limitations are either intrinsic (Montague) or a challenge for compensation.
The lower end of the low-loss area is limited by higher-order resonances.
It is here where the nonlinear multipole components become relevant by shrinking the available tune space.

\begin{figure}[htbp]
    \centering
    \includegraphics[width=0.5\linewidth]{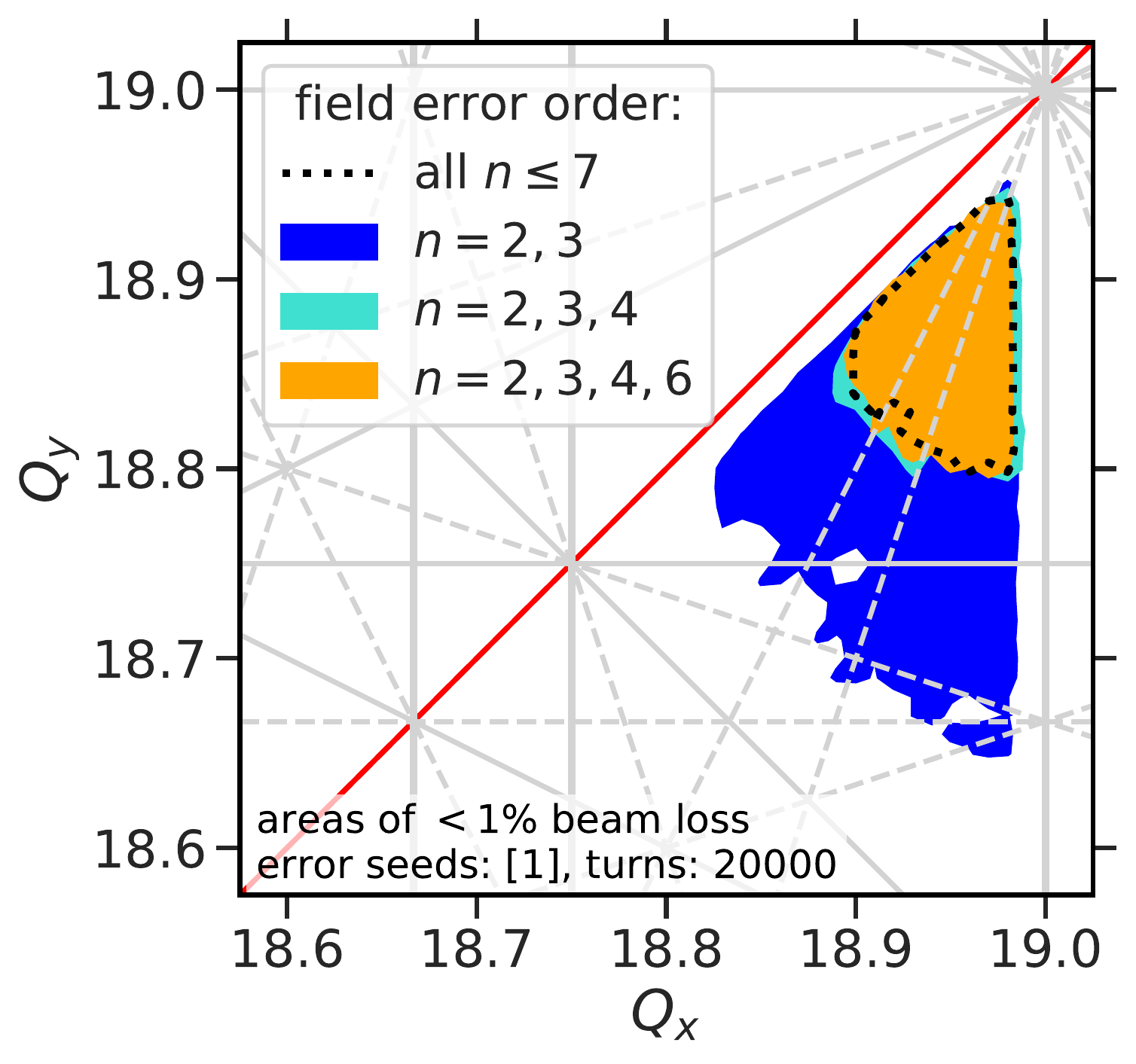}
    \caption{Loss-free tune areas for various included field error orders $n$ (referring to both normal and skew component) in both dipole and quadrupole magnet units (simulations for 20'000 turns).}
    \label{fig: resonance order}
\end{figure}

This is best shown by switching on higher-order multipole components one by one:
the dependence of the loss-free tune area extent on the included multipole order $n$ (both normal $b_n$ and skew $a_n$ are included simultaneously) is displayed in Fig.\ \ref{fig: resonance order}.

The blue area includes the field errors of both dipole and quadrupole magnets up to sextupole order.
Next, the turquoise area includes up to octupole order, which is already almost congruent with the area enclosed by the dashed reference line (which includes all orders of the field error model).
Finally, the orange area just adds the dodecapole $b_6$ component (which on its own without taking into account $n=3,4$ can lead to finite beam loss on the order of $1\%$ along the affiliated resonances) -- $b_6$ explains the small discrepancy between the turquoise area and the black dotted reference line.

In a nutshell, field imperfections of sextupolar and octupolar order are the relevant driving terms which severely limit the available tune space to the bottom.

\subsection{Comments on Results from Ensemble Field Error Model}

The octupole components $b_4$ and $a_4$ are both significantly larger in the FoS 2 model. 
While $b_4$ can drive the normal $4Q_y=75$ resonance located below the lower end of the low-loss working point area, the $a_4$ component can drive the skew $3Q_x-Q_y=38$ resonance which directly crosses the low-loss area.
It thus appears natural to identify the $a_4$ component as the main reason for the larger low-loss area in the ensemble model case.
This hypothesis is tested and confirmed by reintroducing a strong $a_4$ component in the ensemble model:
using the $a_4$ figure from the FoS 2 model while keeping all other $a_n$ and $b_n$ as in the ensemble model, we have run another set of simulations for this tune quadrant.
Indeed, these simulation results predict again a reduced low-loss area of equivalent size as based on the FoS 2 model.

\subsection{Perfectly Linear Quadrupole Units}

\begin{figure}[htbp]
    \centering
    \includegraphics[width=0.5\linewidth]{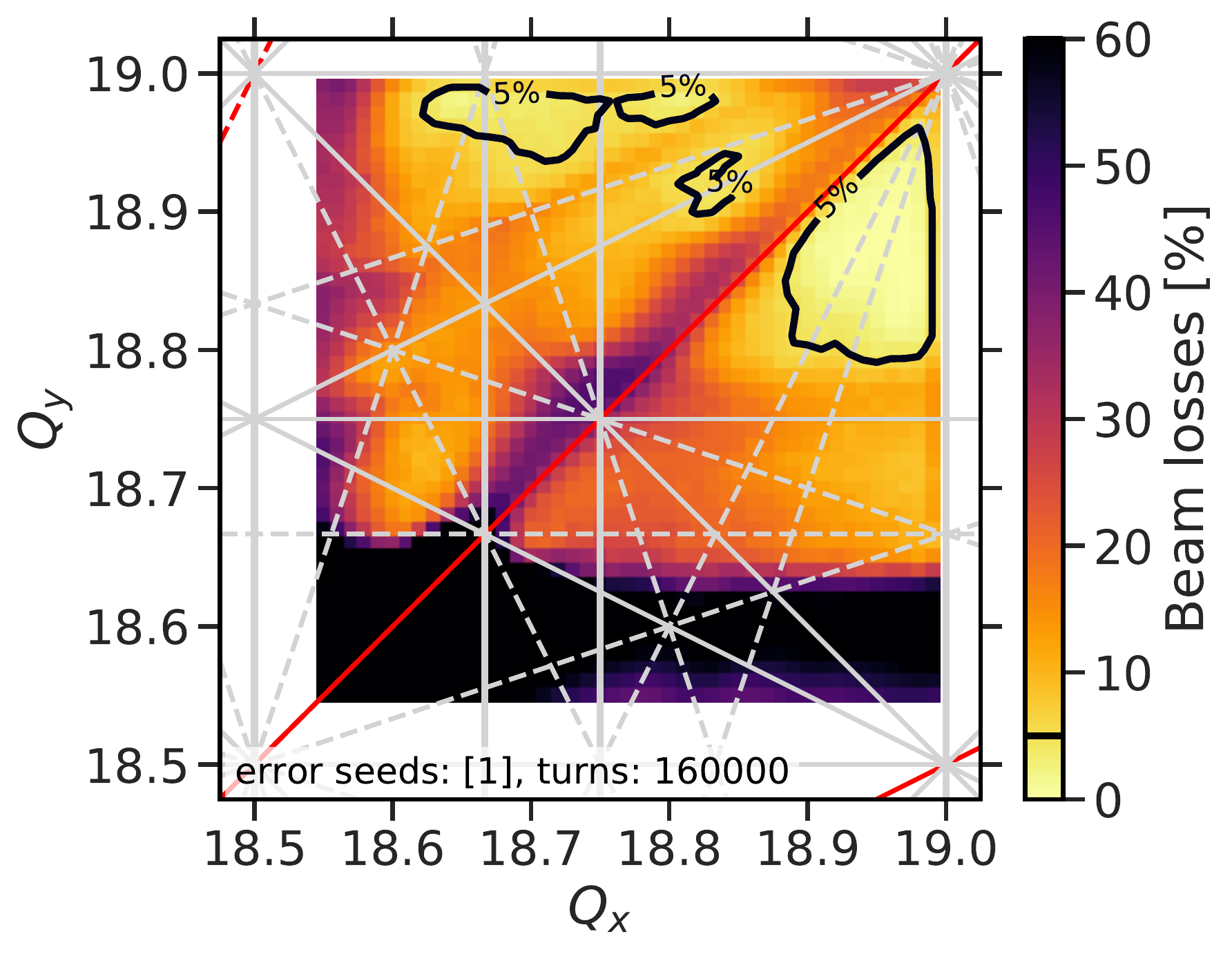}
    \caption{Linear and nonlinear dipole unit field errors and only linear gradient errors in quadrupole units. Beam loss tune diagram during full injection plateau. Resonance lines up to quadrupolar order displayed.} 
    \label{fig: tune diagram nonlinSC diperrs}
\end{figure}

Finally we investigate the scenario of perfectly linear quadrupole units for reference. 
The nonlinear field imperfections of the quadrupole units are set to zero in the error model.
Only the dipole unit field imperfections provide driving terms for higher-order nonlinear resonances.
These simulation results are presented in Fig.\ \ref{fig: tune diagram nonlinSC diperrs}.
The tune diagram offers more low-loss areas.
Without resorting to careful sorting of single magnet units (with the goal to cancel certain resonances), this picture can be understood as an optimum situation with regard to the quadrupole unit quality -- the dipole magnet field quality is now set in stone with its completed unit production.
The previously identified working point region is of similar extent as in the ensemble model case: this confirms satisfying field quality of the quadrupole unit series production w.r.t.\ resonance-induced beam loss during accumulation.

\section{Conclusion}

Beam losses for heavy-ion bunches with the nominal space charge parameters have been simulated for a storage time of \SI{1}{\second} corresponding to the SIS100 injection plateau duration, considering two different scenarios of the nonlinear field error model for the quadrupole units: the First-of-Series \#2 measurement based model (FoS 2) and the ``ensemble'' model based on the recently measured first ten quadrupole units from the ongoing series production. 
The ensemble model results in a larger acceptable low-loss working point area in the tune diagram compared to the FoS 2 model. 
This discrepancy has been traced to the significantly reduced skew octupolar $a_4$ component in the series production quadrupole units.
In general it has been found that the low-loss working point area can be enlarged by minimising sextupolar and octupolar field errors.
All higher multipole orders are already sufficiently weak and affiliated resonances remain relatively harmless across the injection plateau for the studied nominal U${}^{28+}$ intensities.

\printbibliography

\end{document}